\RequirePackage{fix-cm}
\documentclass[smallextended]{svjour3}       % onecolumn (second format)
\smartqed  % flush right qed marks, e.g. at end of proof

\usepackage{graphicx}

\begin{document}

\title{Entangled state fusion with Rydberg atoms}

\author{Y. Q. Ji$^{1,2}$  \and C. M. Dai$^{1,2}$ \and X. Q. Shao$^{1,2}$ \and X. X. Yi$^{1,2,*}$}


\institute{   \at
		      $^*$\email{yixx@nenu.edu.cn}\\
              $^1$ Center for Quantum Sciences and School of Physics, Northeast Normal University, Changchun, 130024, People's Republic of China\\
              $^2$ Center for Advanced Optoelectronic Functional Materials Research, and Key Laboratory for UV Light-Emitting Materials and Technology of Ministry of Education, Northeast Normal University, Changchun 130024, China}

\date{Received: date / Accepted: date}

\maketitle

\begin{abstract}
We propose a scheme for preparation of large-scale entangled $GHZ$ states and $W$ states with neutral Rydberg atoms. The scheme mainly depends on Rydberg antiblockade effect, i.e., as the Rydberg-Rydberg-interaction (RRI) strength and the detuning between the atom transition frequency and the classical laser frequency satisfies some certain conditions, the effective Rabi oscillation between the two ground states and the two excitation Rydberg states would be generated. The prominent advantage is that both two-multiparticle $GHZ$ states and two-multiparticle $W$ states can be fused in
this model, especially the success probability for fusion of $GHZ$ states can reach unit. In addition, the imperfections induced by the spontaneous emission is also discussed through numerical simulation.
\end{abstract}

\section{Introduction}\label{One}
\noindent Due to the promise of applications and rapid experimental progress, the field of quantum information has attracted extensive research. The most advanced experimental demonstrations include trapped ions~\cite{001}, linear optics~\cite{002}, superconductors~\cite{003,004} and quantum dots in semiconductors~\cite{005,006,007}. As a promising candidate for the quantum computer, neutral atom displays another promising approach due to its long-lived encoding of quantum information in atomic hyperfine states and the possibility of manipulating.

When excited to Rydberg state, neutral atom exhibit large dipole moments, which leads to strong and long-range van der Waals or dipole-dipole interactions. The strong and long-range interaction between the excited Rydberg atoms can give rise to the Rydberg blockade that suppress resonant optical excitation of multiple Rydberg atoms~\cite{008,009,010,011}. The Rydberg blockade is based on the assumption that one excited atom causes sufficiently large energy shifts of Rydberg states and leads the neighboring atoms away from resonance with laser field and fully blocks its excitation. It is predicted in ref~\cite{012} and locally observed in laser cooled atomic systems prepared in magneto-optical traps, both for van der Waals~\cite{013,014,015} and dipole-dipole interactions~\cite{009,017}. Recently, the phenomenon of Rydberg blockade with two Rydberg atoms located about 4 $\mu m$~\cite{018} and 10 $\mu m$~\cite{019} away, respectively, have been observed in experiments. The Rydberg blockade offers many possibilities for realization of the neutral-atom-based quantum information processing (QIP) tasks~\cite{Saffman,020,Yang,Khazali} and observation of the multiatom effects~\cite{021,022,023,024,025,026,027}.

When the Rydberg interaction strengths that are too weak to yield the blockade mechanism, yet too strong to be ignored when the atoms are excited with resonant laser fields. The atoms can be excited to the collective Rydberg states~\cite{008}. In addition, when the detuning between the atom transition frequency and the frequency of a classical laser satisfies some conditions with Rydberg interaction strength, the atoms also can be excited to the collective Rydberg states. Hence, the Rydberg antiblockade regime can be generated. This case has theoretically been studied and used for preparation of entanglement and logic gate~\cite{028,029,030,031,032,0320,03200}.

As two of the most basic entangled states, the Greenberger-Horne-Zeilinger ($GHZ$) states~\cite{0321} and the $W$ states~\cite{0322,0323} show great advantage and play an important role in QIP. These two kinds of entangled states can perform different tasks of quantum information theory~\cite{0324,0325,0326}. Therefor, the preparation of entangled state is particularly important. However, it is difficult to create multipartite entangled states in a realistic situation because the dynamics becomes more complex as the number of particles increases. Thus simple and efficient schemes to prepare large-scale multipartite entangled states are very important. In recent works, quantum state fusion has been put forward to realize large-size multipartite entangled states~\cite{0327,0328,0329,0330,0331}, i.e., one can get a larger entangled state from two entangled states on the condition that one qubit of each entangled states is sent to the fusion operation. Such as, \"{O}zdemir \emph{et al}. used a simple optical fusion gate to get an ($m+n-2$)-qubit $W$ states from an $m$-qubit $W$ states and an $n$-qubit $W$ states~\cite{0327}. Nevertheless, most schemes are just for fusion $W$ states.

In this paper, we consider the implementation of entangled states fusion with Rydberg atoms confined in spatially separated dipole traps subject to the Rydberg antiblockade effect. We derive the effective Hamiliton of a two-atoms computational subspace and show how to tailor it in order to implement the specific evolution. Then entangled states fusion with Rydberg atoms can be implemented using this specific evolution. Our scheme has the following characteristics: (1)Adopting neutral atoms as qubits, the quanutm information is encoded into the stable hyperfine ground states and distant atoms interact with each other through the RRI. (2) Only one RRI term
is produced and thus the asymmetric Rydberg coupling strengths are avoided. (3) Not only two multiparticle $W$ states can be fused but also two multiparticle $GHZ$ states can be fused in this model, especially the success probability for fusion of $GHZ$ states can reach unit.

The rest of the paper is organized as follows. In Sec.~\ref{two}, we derive the effective Hamiliton of two Rydberg. In Sec.~\ref{three}
and Sec.~\ref{four}, we describe how to fuse two multiparticle $GHZ$ states and multiparticle $W$ states, respectively.
In Sec.~\ref{five}, a discussion is given. At last, a summary is given in Sec.~\ref{six}.

\section{Basic model}\label{two}
\begin{figure}
\centering\scalebox{0.6}{\includegraphics{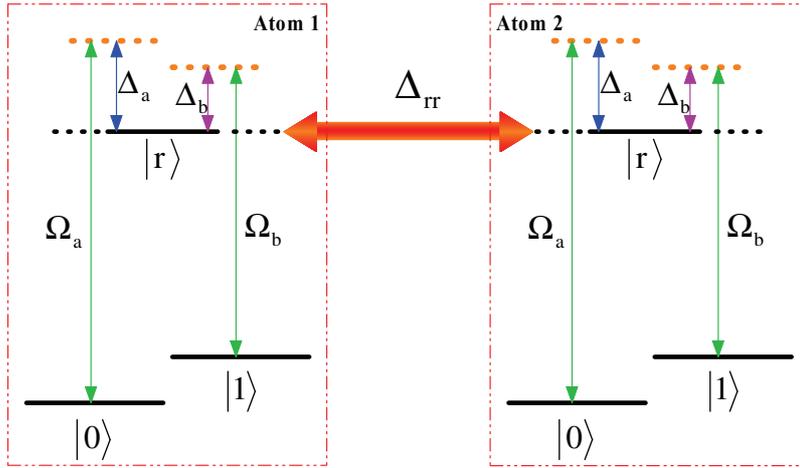}}
\caption{\label{P01}Schematic view of atomic-level configuration. $|r\rangle$ is the
Rydberg state, while $|0\rangle$ and $|1\rangle$ are two ground states. $\Delta_{rr}$ denotes the RRI strength. The atomic transition $|0\rangle\leftrightarrow|r\rangle$ are driven by a classical laser field with the Rabi frequency $\Omega_{a}$ and the transition
$|1\rangle\leftrightarrow|r\rangle$ are driven by a classical laser field with the Rabi frequency $\Omega_{b}$. $\Delta_{a(b)}$ represents the
corresponding detuning parameter.}
\end{figure}
\noindent We consider two identical $^{87}{\rm Rb}$ atoms individually trapped in two tightly focused optical tweezers~\cite{033,034}, with a typical separation of 5-10 microns, and the relevant configuration of atomic level is illustrated in Fig~\ref{P01}, each one has two ground states $|0\rangle$ and $|1\rangle$ and a Rydberg state $|r\rangle$, where $|0\rangle$ and $|1\rangle$ corresponds to atomic levels
$|F=1,M=1\rangle$ and $|F=2,M=2\rangle$ of $5S_{1/2}$ manifold, and the Rydberg state $|r\rangle=|F=3,M=3\rangle$ of $58D_{3/2}$
directly coupled to the ground states by a single exciting laser. The transition $|0(1)\rangle\leftrightarrow|r\rangle$ is then driven by a nonresonant classical laser field with Rabi frequencies $\Omega_{a(b)}$, frequency $\omega_{a(b)}$, and detuning $\Delta_{a(b)}$.  As is well known, Rydberg atoms exhibit huge dipole moments which lead to large dipole-dipole interactions. We shall assume that the dipole-dipole interaction manifests itself only when both of atoms are in their Rydberg states. In other words, we assume $\hat{U}_{rr}=\Delta_{rr}|rr\rangle\langle rr|$ and $\Delta_{rr}$ is the RRI strength which mainly depends on the principal quantum numbers of the Rydberg atoms and the distance between the Rydberg atoms. Thus, the Hamiltonian of the whole system can be written as
\begin{eqnarray}\label{01}
\hat{H}=\hat{H_{1}}\otimes{\hat{I}}_2+{\hat{I}}_1\otimes \hat{H_{2}}+\hat{U}_{rr},
\end{eqnarray}
with
\begin{eqnarray}\label{02}
\hat{H}_{p}&=&\frac{\Omega_{a}}{2}e^{i\omega_{a}t}|0\rangle_{p}\langle r|+\frac{\Omega_{b}}{2}e^{i\omega_{b}t}|1\rangle_{p}\langle r|+\rm{H.c.} \cr
&&+\sum_{j=0,1,r}\omega_{j}|j\rangle_{p}\langle j|,
\end{eqnarray}
${\hat{I}}_p$ denotes the $3 \times 3$ identity matrix ($p=1, 2$). After moving $\hat{H}$ to the interaction picture with respect to
Hamiltonian $\sum_{p=1,2}\sum_{j=0,1,r}\omega_{j}|j\rangle_{p}\langle j|$, we can get
\begin{eqnarray}\label{03}
\hat{H}_{p}'=\frac{\Omega_{a}}{2}e^{i\Delta_{a}t}|0\rangle_{p}\langle r|+\frac{\Omega_{b}}{2}e^{i\Delta_{b}t}|1\rangle_{p}\langle r|+\rm{H.c.},
\end{eqnarray}
and $\hat{U}_{rr}$ remains unchanged. To see clearly the role of the RRI term, we rewrite the full Hamiltonian with the basis $\{|00\rangle,|01\rangle,|0r\rangle,|10\rangle,|11\rangle,|1r\rangle,|r0\rangle,|r1\rangle,|rr\rangle\}$ and move to the rotation frame with respect to $\hat{U}_{rr}$. Thus, the full Hamiltonian is transformed to
\begin{eqnarray}\label{04}
\hat{H}''&=&\frac{\Omega_{a}}{2}e^{i\Delta_{a}t}(|00\rangle\langle r0|+|10\rangle\langle rb|+|00\rangle\langle 0r|+|01\rangle\langle 1r|) \cr
&&+\frac{\Omega_{a}}{2}e^{i(\Delta_{a}-\Delta_{rr})t}(|r0\rangle\langle rr|+|0r\rangle\langle rr|) \cr
&&+\frac{\Omega_{b}}{2}e^{i\Delta_{b}t}|01\rangle\langle r0|+|11\rangle\langle r1|+|10\rangle\langle 0r|+|11\rangle\langle 1r|) \cr
&&+\frac{\Omega_{b}}{2}e^{i(\Delta_{b}-\Delta_{rr})t}(|r1\rangle\langle rr|+|1r\rangle\langle rr|)+\rm{H.c.}.
\end{eqnarray}
In the following, we assume $\Omega_{a}=\Omega_{b}=\Omega$ and $\Delta_{a}=\Delta_{b}=\Delta$, $\Delta_{a}=\omega_{a}-(\omega_{r}-\omega_{0})$ and $\Delta_{b}=\omega_{b}-(\omega_{r}-\omega_{1})$. Note that, whether $\omega_{0}$ and $\omega_{1}$ are equal depend on $\omega_{a}$ and $\omega_{b}$. In fact, $\omega_{a}$ and $\omega_{b}$ are not equal. So the states $|0\rangle$ and $|1\rangle$ are not degenerate. To achieve the antiblockade regime, we here adjust the classical field and RRI strength to make the parameters satisfy $2\Delta=\Delta_{rr}$. Then, the large detuned condition
$\Delta\gg\Omega/2$ would induce the effective Hamiltonian~\cite{035}
\begin{small}
\begin{eqnarray}\label{05}
\hat{H}_{e}&=&\frac{\Omega^2}{2\Delta}\left[(|00\rangle+|rr\rangle)(\langle 00|+\langle rr|)+(|11\rangle+|rr\rangle)(\langle 11|+\langle rr|)\right] \cr
&&-\frac{\Omega^2}{2\Delta}\left[(|0r\rangle+|1r\rangle)(\langle 0r|+\langle 1r|)+(|r0\rangle+|r1\rangle)(\langle r0|+\langle r1|)\right] \cr
&&+\frac{\Omega^2}{4\Delta}\left[(|00\rangle+|11\rangle)(\langle 10|+\langle 01|)+(|01\rangle+|10\rangle)(\langle 00|+\langle 11|)\right] \cr
&&-\frac{\Omega^2}{4\Delta}\left[(|0r\rangle+|1r\rangle)(\langle r0|+\langle r1|)+(|r0\rangle+|r1\rangle)(\langle 0r|+\langle 1r|)\right] \cr
&&-\frac{\Omega^2}{2\Delta}\left[|0r\rangle\langle r0|+|1r\rangle\langle r1|+|r0\rangle\langle 0r|+|r1\rangle\langle 1r|\right] \cr
&&+\frac{\Omega^2}{2\Delta}\left[(|01\rangle+|10\rangle)\langle rr|+|rr\rangle(\langle 01|+\langle 10|)\right] \cr
&&+\frac{\Omega^2}{2\Delta}(|01\rangle\langle 10|+|10\rangle\langle 01|).
\end{eqnarray}
\end{small}
Because the initial state is among the basis $\{|00\rangle,|01\rangle,|10\rangle,|11\rangle\}$, the states $|0r\rangle$, $|r0\rangle$, $|1r\rangle$ and $|r1\rangle$ have no energy exchange with the other states, after we reject these states, the Eq.~(\ref{05}) becomes
\begin{small}
\begin{eqnarray}\label{050}
\hat{H}_{\rm eff}&=&\frac{\Omega^2}{2\Delta}\left[(|00\rangle+|rr\rangle)(\langle 00|+\langle rr|)+(|11\rangle+|rr\rangle)(\langle 11|+\langle rr|)\right] \cr
&&+\frac{\Omega^2}{4\Delta}\left[(|00\rangle+|11\rangle)(\langle 01|+\langle 10|)+(|01\rangle+|10\rangle)(\langle 00|+\langle 11|)\right] \cr
&&+\frac{\Omega^2}{2\Delta}\left[(|01\rangle+|10\rangle)\langle rr|+|rr\rangle(\langle 01|+\langle 10|)\right] \cr
&&+\frac{\Omega^2}{2\Delta}(|01\rangle\langle 10|+|10\rangle\langle 01|).
\end{eqnarray}
\end{small}
From Eq.~(\ref{050}) we can see that the effective Rabi oscillation between the two ground states and the two excited Rydberg states are generated, which is out of the Rydberg blockade regime.

For the initial states $|00\rangle$, $|01\rangle$, $|10\rangle$ and $|11\rangle$, the roles of the effective evolution operator
$e^{-i\hat{H}_{\rm eff}t}$ can be illustrated as follows:
\begin{eqnarray}\label{06}
|00\rangle&\rightarrow& a|00\rangle+b|01\rangle+b|10\rangle+c|11\rangle+d|rr\rangle \cr
|01\rangle&\rightarrow& b|00\rangle+a|01\rangle+c|10\rangle+b|11\rangle+d|rr\rangle \cr
|10\rangle&\rightarrow& b|00\rangle+c|01\rangle+a|10\rangle+b|11\rangle+d|rr\rangle \cr
|11\rangle&\rightarrow& c|00\rangle+b|01\rangle+b|10\rangle+a|11\rangle+d|rr\rangle,
\end{eqnarray}
with
\begin{eqnarray}\label{07}
a&=&\frac{1}{8}\left(3+4e^{-i\frac{\Omega^2t}{2\Delta}}+e^{-i\frac{2\Omega^2t}{\Delta}}\right) \cr
b&=&\frac{1}{8}\left(-1+e^{-i\frac{2\Omega^2t}{\Delta}}\right) \cr
c&=&\frac{1}{8}\left(3-4e^{-i\frac{\Omega^2t}{2\Delta}}+e^{-i\frac{2\Omega^2t}{\Delta}}\right) \cr
d&=&\frac{1}{4}\left(-1+e^{-i\frac{2\Omega^2t}{\Delta}}\right).
\end{eqnarray}
After choosing the parameters satisfy $\Omega^2t/\Delta=\pi$, Eq.~(\ref{06}) can be simplified to
\begin{eqnarray}\label{08}
|00\rangle&\rightarrow&\frac{1}{\sqrt{2}}\left(|00\rangle+i|11\rangle\right) \cr
|01\rangle&\rightarrow&\frac{1}{\sqrt{2}}\left(|01\rangle+i|10\rangle\right) \cr
|10\rangle&\rightarrow&\frac{1}{\sqrt{2}}\left(|10\rangle+i|01\rangle\right) \cr
|11\rangle&\rightarrow&\frac{1}{\sqrt{2}}\left(|11\rangle+i|00\rangle\right),
\end{eqnarray}
in this process, we ignore the global phase factor $e^{-i\frac{\pi}{4}}$.

For the result of Eq.~(\ref{08}), one can find some interesting things, such as preparation of entangle states, i.e., the entanglement between
$|00\rangle$ and $|11\rangle$ or $|01\rangle$ and $|10\rangle$.

\section{Fusing atomic $GHZ$ states}\label{three}
\noindent Now, we introduce how to implement an ($m+n-2$)-qubit $GHZ$ states fusion scheme from an $m$-qubits $GHZ$ states and an $n$-qubits $GHZ$ states based on Rydberg atoms, where $m\geq3$ and $n\geq3$. The entangled $GHZ$ states of Alice and Bob are
\begin{eqnarray}\label{10}
|GHZ_{m}\rangle_{A}&=&\frac{1}{\sqrt{2}}\left(|(m-1)_{0}\rangle|1_{0}\rangle+|(m-1)_{1}\rangle|1_{1}\rangle\right),\cr
|GHZ_{n}\rangle_{B}&=&\frac{1}{\sqrt{2}}\left(|(n-1)_{0}\rangle|1_{0}\rangle+|(n-1)_{1}\rangle|1_{1}\rangle\right).
\end{eqnarray}
Here $|(m-1)_{0}\rangle$ denotes the ($m-1$) atoms remain $|0\rangle$. To start the fusion process, the two atoms, respectively, belong to Alice and Bob, will be sent into the third party Claire who receives two atoms with Rydberg antiblockade effect to merge and inform
them when the task is successful. So the initial state of the whole system is
\begin{eqnarray}\label{11}
|\psi_{0}\rangle=|GHZ_{m}\rangle_{A}\otimes|GHZ_{n}\rangle_{B}.
\end{eqnarray}
The far-off-resonant interaction between the classical laser field and the two atoms will lead the initial states evolve to the
following state(according the result in Eq.~(\ref{06}))
\begin{eqnarray}\label{12}
|\psi_{1}\rangle&=&\frac{1}{2}|(m-1)_{0}\rangle|(n-1)_{0}\rangle\frac{1}{\sqrt{2}}\left(|00\rangle+i|11\rangle\right) \cr
&+&\frac{1}{2}|(m-1)_{0}\rangle|(n-1)_{1}\rangle\frac{1}{\sqrt{2}}\left(|01\rangle+i|10\rangle\right) \cr
&+&\frac{1}{2}|(m-1)_{1}\rangle|(n-1)_{0}\rangle\frac{1}{\sqrt{2}}\left(|10\rangle+i|01\rangle\right) \cr
&+&\frac{1}{2}|(m-1)_{1}\rangle|(n-1)_{1}\rangle\frac{1}{\sqrt{2}}\left(|11\rangle+i|00\rangle\right).
\end{eqnarray}
Then the two atoms will be detected by Claire. There are four possible detection results for the fusion mechanism. If the detection result is $|00\rangle$, the remaining atoms are in the following state
\begin{eqnarray}\label{13}
|\psi_{2}\rangle=\frac{1}{\sqrt{2}}(|(m-1)_{0}\rangle|(n-1)_{0}\rangle+i|(m-1)_{1}\rangle|(n-1)_{1}\rangle).
\end{eqnarray}
There are only relative phase differences between the state $|\psi_{2}\rangle$ and the standard $GHZ$ states. After Alice or Bob performs the one-qubit phase gate on one of the atoms that she or he has, i.e.,
\{$|0\rangle\rightarrow |0\rangle,|1\rangle\rightarrow i|1\rangle$\}, the state in Eq.~(\ref{13}) will become an $m+n-2$-qubit $GHZ$ states and the corresponding success probability is 1/4.

If the detection result is $|11\rangle$, the systemic state will collapse to
\begin{eqnarray}\label{14}
|\psi_{3}\rangle=\frac{1}{\sqrt{2}}(i|(m-1)_{0}\rangle|(n-1)_{0}\rangle+|(m-1)_{1}\rangle|(n-1)_{1}\rangle),
\end{eqnarray}
similarly, we also can obtain an $(m+n-2)$-qubit $GHZ$ states if Alice or Bob performs similar operation.

If the detection result is $|01\rangle$ or $|10\rangle$, the systemic state will collapse to
\begin{eqnarray}\label{15}
|\psi_{4}\rangle=\frac{1}{\sqrt{2}}(|(m-1)_{0}\rangle|(n-1)_{1}\rangle+i|(m-1)_{1}\rangle|(n-1)_{0}\rangle)
\end{eqnarray}
or
\begin{eqnarray}\label{16}
|\psi_{5}\rangle=\frac{1}{\sqrt{2}}(i|(m-1)_{0}\rangle|(n-1)_{1}\rangle+|(m-1)_{1}\rangle|(n-1)_{0}\rangle),
\end{eqnarray}
respectively. Also the states in Eqs~(\ref{15},\ref{16}) can be transformed into a standard $GHZ$ states by one-qubit phase gate on any one
of the ($m+n-2$) atoms. Hence, the total success probability for fusion an $m$-qubit $GHZ$ states and an $n$-qubit $GHZ$ state can reach unit.

\section{Fusing atomic $W$ states}\label{four}
\noindent Now, we introduce how to implement an ($m+n-2$)-qubit atomic $W$ states fusion scheme. The atomic entangled $W$ states of Alice and Bob are
\begin{eqnarray}\label{17}
|W_{m}\rangle_{A}&=&\frac{1}{\sqrt{m}}\left(|(m-1)_{0}\rangle|1_{1}\rangle+\sqrt{m-1}|W_{m-1}\rangle|1_{0}\rangle\right),\cr
|W_{n}\rangle_{B}&=&\frac{1}{\sqrt{n}}\left(|(n-1)_{0}\rangle|1_{1}\rangle+\sqrt{n-1}|W_{n-1}\rangle|1_{0}\rangle\right).
\end{eqnarray}
To start the fusion process, the two atoms (atom1 and atom 2) will be sent into the third party Claire. So the initial state of the whole system is
\begin{eqnarray}\label{18}
|\phi_{0}\rangle=|W_{m}\rangle_{A}\otimes|W_{n}\rangle_{B}.
\end{eqnarray}
The far-off-resonant interaction between the classical laser field and the two atoms will lead the initial states evolve to the
following state
\begin{eqnarray}\label{19}
|\phi_{1}\rangle&=&\frac{1}{\sqrt{mn}}|(m-1)_{0}\rangle|(n-1)_{0}\rangle\otimes\frac{1}{\sqrt{2}}(|11\rangle+i|00\rangle) \cr
&&+\frac{\sqrt{(m-1)(n-1)}}{\sqrt{mn}}|W_{m-1}\rangle|W_{n-1}\rangle\otimes\frac{1}{\sqrt{2}}(|00\rangle+i|11\rangle)\cr
&&+\frac{\sqrt{n-1}}{\sqrt{mn}}|(m-1)_{0}\rangle|W_{n-1}\rangle\otimes\frac{1}{\sqrt{2}}(|10\rangle+i|01\rangle)\cr
&&+\frac{\sqrt{m-1}}{\sqrt{mn}}|W_{m-1}\rangle|(n-1)_{0}\rangle\otimes\frac{1}{\sqrt{2}}(|01\rangle+i|10\rangle).
\end{eqnarray}

Then Claire will detect the states of two atoms and inform Alice and Bob whether the task is successful.
There are four possible detection results from Eq.~(\ref{19}).
If the detection result is $|10\rangle$, the remaining atoms are in the following state
\begin{small}
\begin{eqnarray}\label{20}
|\phi_{2}\rangle=\frac{1}{\sqrt{2mn}}\left(\sqrt{n-1}|(m-1)_{0}\rangle|W_{n-1}\rangle+\sqrt{m-1}i|W_{m-1}\rangle|(n-1)_{0}\rangle\right).
\end{eqnarray}
\end{small}
After Bob performs the one-qubit phase gate on all the atoms that he has, i.e., \{$|0\rangle\rightarrow |0\rangle,|1\rangle\rightarrow i|1\rangle$\},
the state in Eqs.~(\ref{20}) will become
\begin{small}
\begin{eqnarray}\label{21}
|\phi'_{2}\rangle&=&\frac{1}{\sqrt{2mn}}\left(\sqrt{n-1}|(m-1)_{0}\rangle|W_{n-1}\rangle+\sqrt{m-1}|W_{m-1}\rangle|(n-1)_{0}\rangle\right)\cr
&=&\frac{\sqrt{n+m-2}}{\sqrt{2mn}}|W_{m+n-2}\rangle,
\end{eqnarray}
\end{small}
where we have ignored the global phase factor $i$ and used $\sqrt{k}|W_{k}\rangle$=$\sqrt{i}|W_{i}\rangle|(k-i)_{g_{0}}\rangle$+$\sqrt{i-1}|i_{g_{0}}\rangle|W_{k-i}\rangle$.
Obviously, $|\phi'_{2}\rangle$ is a standard atomic $W$ states, i.e., $|W_{n+m-2}\rangle$, and the success probability obtaining the $|\phi'_{2}\rangle$ state is $(n+m-2)/(2mn)$.

If the detection result is $|01\rangle$, the systemic state becomes
\begin{small}
\begin{eqnarray}\label{22}
|\phi_{3}\rangle=\frac{1}{\sqrt{2mn}}\left(i\sqrt{n-1}|(m-1)_{0}\rangle|W_{n-1}\rangle+\sqrt{m-1}|W_{m-1}\rangle|(n-1)_{0}\rangle\right).
\end{eqnarray}
\end{small}
After Alice performs the one-qubit phase gate on all the atoms that she has, the state in Eq.~(\ref{22}) will become Eq.~(\ref{21}),
and the success probability is $(n+m-2)/(2mn)$. However, the cases of $|00\rangle$ and $|11\rangle$ are failure. Thus the total success probability for the fusion process is
\begin{eqnarray}\label{23}
P_{n+m-2}=\frac{n+m-2}{mn}.
\end{eqnarray}

\section{Discussion}\label{five}
\begin{figure}
\centering\scalebox{0.6}{\includegraphics{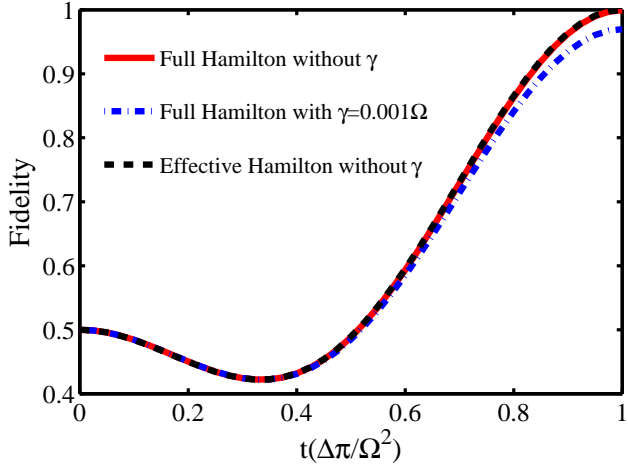}}
\caption{\label{P1}Choosing the initial state $|00\rangle$ and final state $\frac{1}{\sqrt{2}}(|00\rangle+i|11\rangle)$ at
the time interval $t\in [0,\Delta\pi/\Omega^{2}]$ to check the performance. The red solid line and black dashed line, respectively,
denotes the fidelity with the full Hamilton and effective Hamilton with out spontaneous emission $\gamma$. The blue dot-dashed line
denotes the fidelity with the full Hamilton with $\gamma=0.001\Omega$.
The other parameters are chosen as $\Omega_{a}=\Omega_{b}=\Omega=1$, $\Delta_{a}=\Delta_{b}=\Delta=40$, $\Delta_{rr}=2\Delta=80$.}
\end{figure}
\begin{figure}
\centering\scalebox{0.6}{\includegraphics{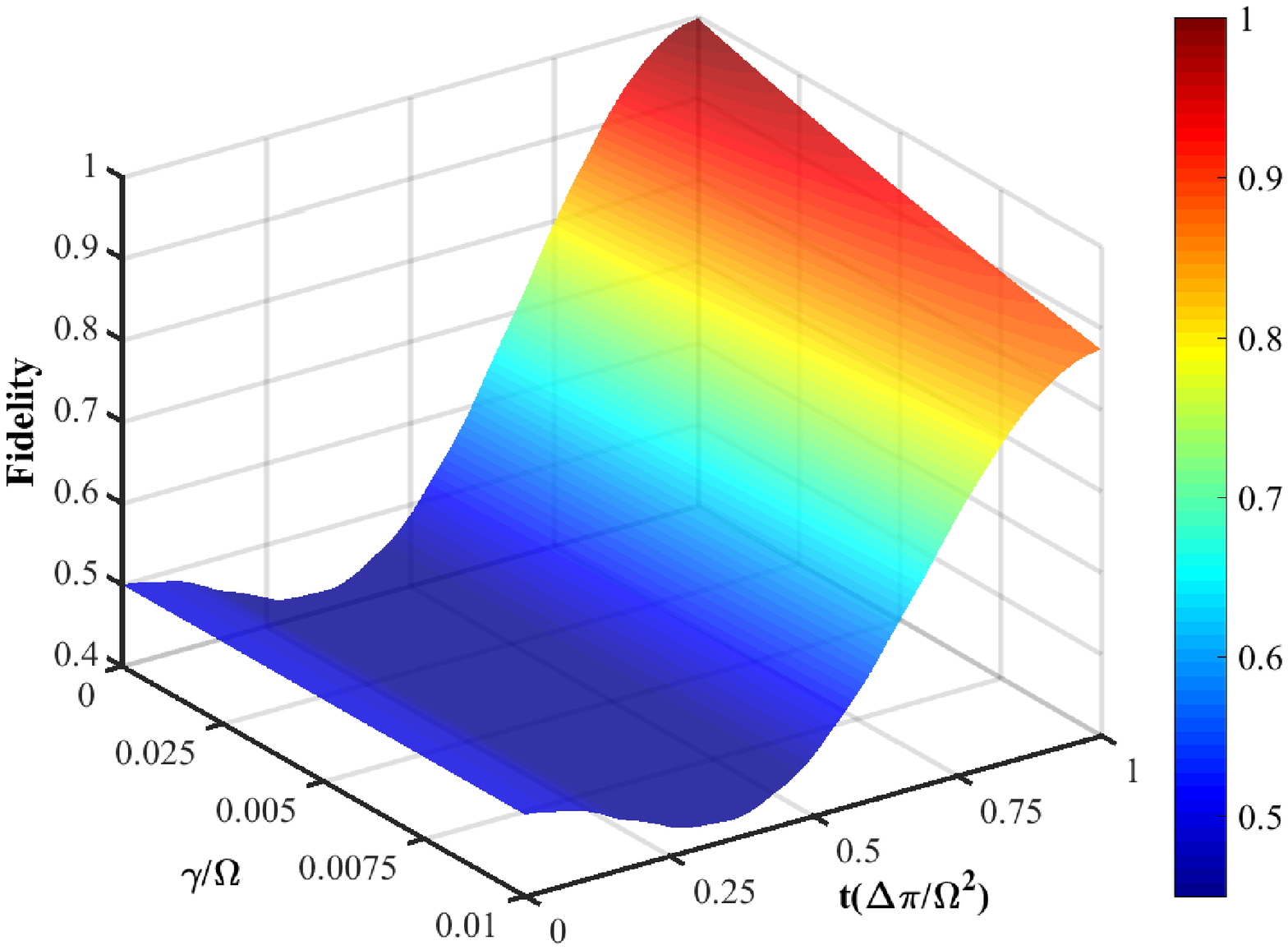}}
\caption{\label{P2}The effect of decoherence induced by spontaneous emission of Rydberg state as well as evolution time $t$ for the evolution
of initial states $|00\rangle$. The parameters are chosen as $\Omega_{a}=\Omega_{b}=\Omega=1$, $\Delta_{a}=\Delta_{b}=\Delta=40$, $\Delta_{rr}=2\Delta=80$.}
\end{figure}

\noindent In the above, the dissipation has not been taken into account. Thus, we investigate the influence of spontaneous emission on this method. When decoherence effects are taken into account and under the assumptions that the decay channels are independent, the master equation of the whole system can be expressed by the Lindblad form~\cite{036,037}
\begin{eqnarray}\label{24}
\dot{\rho}=-i[H,\rho]
-\frac{1}{2}\sum_{k=1}^{4}\left[\hat{\mathcal{L}}_{k}^{\dag}\hat{\mathcal{L}_{k}}\rho-2\hat{\mathcal{L}_{k}}\rho\hat{\mathcal{L}_{k}^{\dag}}
+\rho\hat{\mathcal{L}_{k}^{\dag}}\hat{\mathcal{L}_{k}}\right],
\end{eqnarray}
where $\rho$ is the density matrix of the whole system and $\gamma$ denotes the spontaneous emission rate,
$\hat{\mathcal{L}_{1}}=\sqrt{\gamma/2}|0\rangle_{1}\langle r|$,
$\hat{\mathcal{L}_{2}}=\sqrt{\gamma/2}|1\rangle_{1}\langle r|$, $\hat{\mathcal{L}_{3}}=\sqrt{\gamma/2}|0\rangle_{2}\langle r|$ and $\hat{\mathcal{L}_{1}}=\sqrt{\gamma/2}|1\rangle_{2}\langle r|$ are Lindblad operators that describe the dissipative processes. For simplicity, here we have assumed the Rydberg state $|r\rangle$ has an equal spontaneous emission rate for the two ground states $|0\rangle$ and $|1\rangle$.

To check the performance, the fidelity is defined as $\langle \psi_{ideal}|\hat\rho(t)|\psi_{ideal}\rangle$.
In Fig.~\ref{P1}, we choose $|00\rangle$ act as initial state and $\frac{1}{\sqrt{2}}(|00\rangle+i|11\rangle)$ act as final state.
One can see that the curves plotted with the full Hamiltonian(without $\gamma$) and effective Hamiltonian(without $\gamma$),
respectively, fit well with each other, which proves the effective Hamiltonian is valid in this paper. In Fig.~\ref{P2}, we use the initial state $|00\rangle$ corresponding final state $\frac{1}{\sqrt{2}}(|00\rangle+i|11\rangle)$ to check the performance. From Fig.~\ref{P2}, we can see that the fidelity is very high at the optimal time and the spontaneous emission lead to the fidelity decrease. For the initial states $|01\rangle$, $|10\rangle$ and $|11\rangle$, decoherence effects is the same to each other, this is because the evolution of $|00\rangle$ to $\frac{1}{\sqrt{2}}(|00\rangle+i|11\rangle)$ must through the intermediate state $|rr\rangle$, the states $|0r\rangle$, $|r0\rangle$, $|1r\rangle$ and $|r1\rangle$ are not exist due to the large detuning. From another perspective, if we change
$|0\rangle$ to $|1\rangle$ or $|1\rangle$ to $|0\rangle$ on each of two atoms, the effective Hamilton remain unchanged. Therefor, the decoherence effects for four final states are considered the same.

\begin{figure}
\centering\scalebox{0.6}{\includegraphics{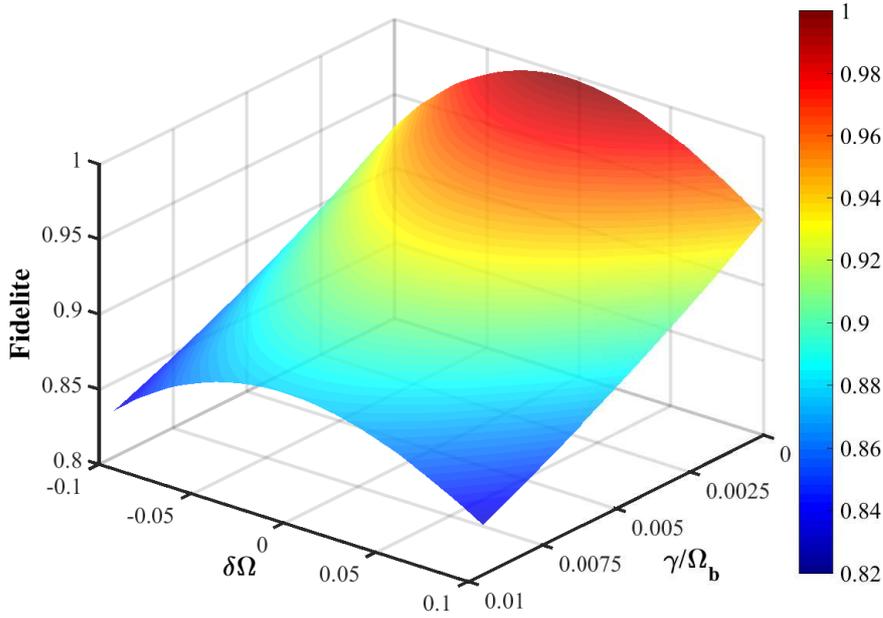}}
\caption{\label{P4}Fidelity with respect to $\gamma/\Omega_{b}$ as well as $\delta\Omega$ ($\delta\Omega=\frac{\Omega_{a}-\Omega_{b}}{\Omega_{b}}$) at
the optimal time $\frac{\Delta\pi}{\Omega^{2}}$. The other parameters are chosen as $\Omega_{b}=1$, $\Delta_{a}=\Delta_{b}=\Delta=40$, $\Delta_{rr}=2\Delta=80$.}
\end{figure}

\begin{figure}
\centering\scalebox{0.6}{\includegraphics{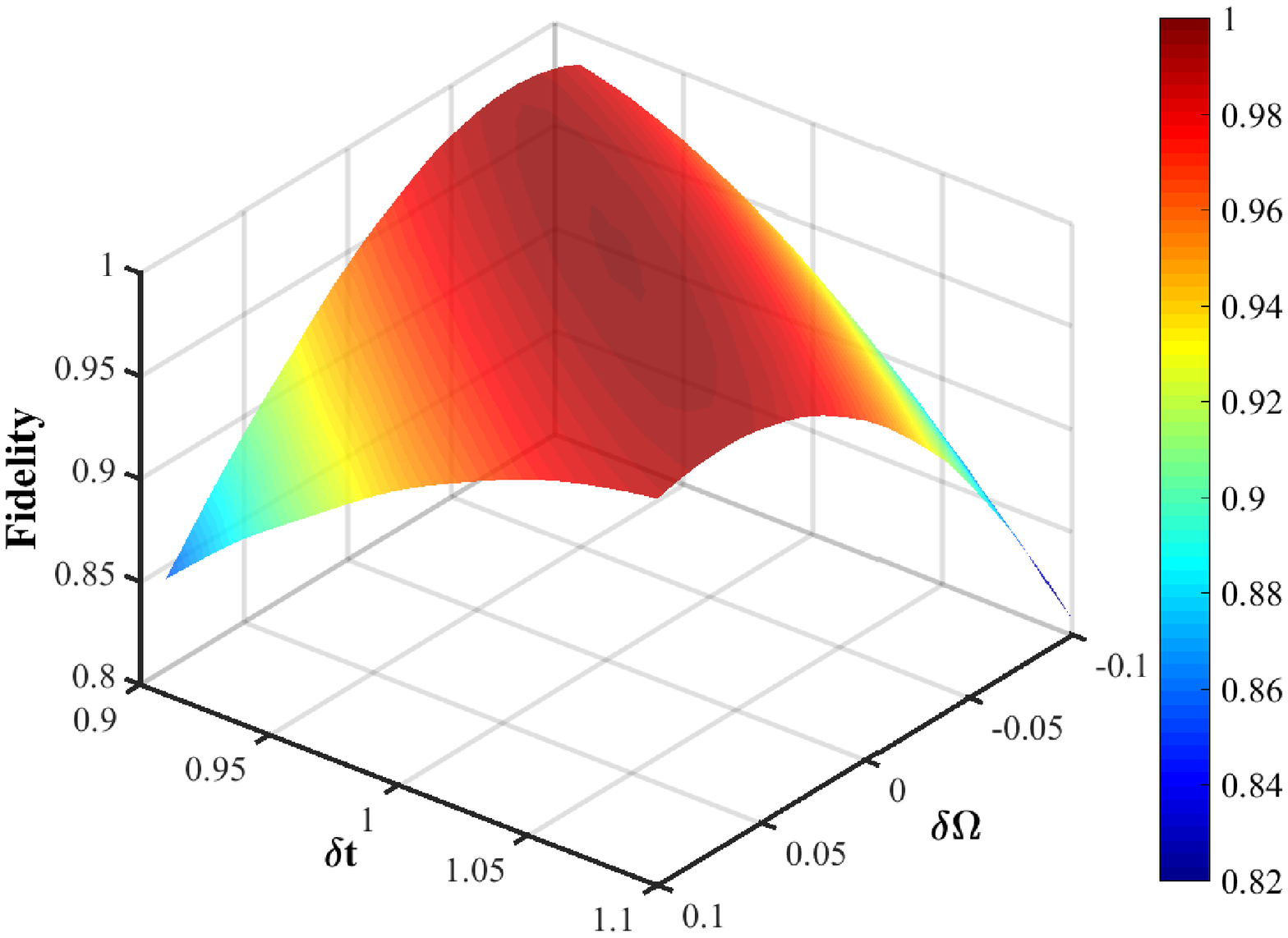}}
\caption{\label{P5}Fidelity with respect to $\delta t$ as well as $\delta\Omega$, where $\delta t=\frac{t}{t_{0}}$,
and $\delta\Omega=\frac{\Omega_{a}-\Omega_{b}}{\Omega_{b}}$, $t$ is actual time of evolution and $t_{0}$ is the optimal evolution time.
The other parameters are chosen as $\Omega_{b}=1$, $\Delta_{a}=\Delta_{b}=\Delta=40$, $\Delta_{rr}=2\Delta=80$ and $\gamma=0$.}
\end{figure}

For the $GHZ$ states fusion, due to the fidelity of the states $\frac{1}{\sqrt{2}}(|00\rangle+i|11\rangle)$,
$\frac{1}{\sqrt{2}}(|01\rangle+i|10\rangle)$, $\frac{1}{\sqrt{2}}(|10\rangle+i|01\rangle)$ and
$\frac{1}{\sqrt{2}}(|11\rangle+i|00\rangle)$ are considered equal to each other, so we assume it equal to $F'$. Hence, the final fidelity
(Eq.~(\ref{12}) as final state) can be represented as $F_{GHZ}=\frac{1}{2}F'+\frac{1}{2}F'+\frac{1}{2}F'+\frac{1}{2}F')=F'$.
However, for the $W$ states fusion, the final fidelity (Eq.~(\ref{19}) as final state)
can be represented as $F_{W}=\frac{1}{mn}F'+\frac{(m-1)(n-1)}{mn}F'+\frac{n-1}{mn}F'+\frac{m-1}{mn}F'=F'$. In the fusion process, single-qubit gates imperfections can be quite small, leading to fidelity errors $O(10^{-4})$~\cite{030}. Hence, we have ignored the influence of one-qubit phase gate in this schemes, i.e., $F_{GHZ}=F_{W}\simeq F'$. The robustness against operational imperfection is also a main factor for the feasibility of the schemes.
In the above numerical simulations, for simplicity we only consider the case $\Omega_{a}=\Omega_{b}=\Omega$. So a numerical simulation
is performed to check the fidelity by varying error parameters of the mismatch among the Rabi frequency of classical laser field and the
interaction time through solving the master equation numerically with the full Hamiltonian. We define $\delta\Omega=\frac{\Omega_{a}-\Omega_{b}}{\Omega_{b}}$
and $\delta t=\frac{t}{t_{0}}$, $t_{0}$ is the optimal evolution time. The fidelity varies the variations in different parameters are shown
in Fig.~\ref{P4} and Fig.~\ref{P5}. In Fig.~\ref{P4}, we plot the fidelity with respect to $\gamma/\Omega_{b}$ as well as $\delta\Omega$ at
the optimal time $\frac{\Delta\pi}{\Omega^{2}}$, where we have set $\Omega_{b}=1$, $\Delta_{a}=\Delta_{b}=\Delta=40$, $\Delta_{rr}=2\Delta=80$.
In Fig.~\ref{P5}, we plot the fidelity with respect to $\delta t$ as well as $\delta\Omega$ when $\Omega_{b}=1$, $\Delta_{a}=\Delta_{b}=\Delta=40$,
$\Delta_{rr}=2\Delta=80$ and $\gamma=0$.
As shown in the figures, the schemes is a little sensitive to the variation in the laser Rabi frequency
and the interaction time. But that is not a serious problem to realize the schemes because the laser Rabi frequency and the interaction
time can be precisely controlled in experiment.
In Ref.~\cite{033,034}, the parameters are chosen as follows: RRI strength $\Delta_{rr}\simeq2\times10^{3}$ MHz, Rydberg state decay rate
$\gamma=10$ kHz, it is reasonable if we set $\Delta_{a}=\Delta_{b}=10^{3}$ MHz, and $\Omega_{a}=\Omega_{b}=50$ MHz. By substituting these
values into the master equation, the fidelities $F'=99.4\%$ for fusion of $GHZ$ states and $W$ states
could be achieved.

\section{Summary}\label{six}
\noindent In summary, we have proposed a method to fuse entangled $GHZ$ states and $W$ states based on neutral Rydberg atoms. This scheme works well in the regime where the Rydberg interaction holds a comparable strength to the detuning.
Final numerical simulation based on one group of experiment parameters shows that our scheme could be feasible under current technology and have a high fidelity. We believe our work will be useful for the experimental realization of quantum information with neutral atoms in the near future.

\section*{ACKNOWLEDGMENTS}
This work is supported by National Natural Science
Foundation of China (NSFC) under Grants No. 11534002, No. 61475033
and Fundamental Research Funds for the
Central Universities under Grant No. 2412016KJ004.

\end{document}